\documentclass[aip,amsmath,amssymb,reprint]{revtex4-2}

\usepackage{graphicx}
\usepackage{tikz}
\usepackage{dcolumn}
\usepackage{bm}
\usepackage{float}
\usepackage{color}
\usepackage{xcolor}
\usepackage{amsmath}

\usepackage{amssymb}
\usepackage{amsbsy}
\usepackage{soul}
\usepackage{titlesec}
\usepackage[nottoc]{tocbibind}
\usepackage[normalem]{ulem}

\usepackage[utf8]{inputenc}
\usepackage[T1]{fontenc}
\usepackage{mathptmx}

\usepackage{color}
\usepackage{soul}

\begin{document}


\title{Dynamic Triad Interactions and Evolving Turbulence - Part 2 Data: Implications for Practical Signals} 



\author{Preben Buchhave}
\email[]{buchhavepreben@gmail.com}
\affiliation{Intarsia Optics, S{\o}nderskovvej 3, 3460 Birker{\o}d, Denmark}

\author{Clara M. Velte}
\email[]{cmve@dtu.dk}
\homepage[Personal web page: ]{https://www.staff.dtu.dk/cmve}
\homepage[Projects web page: ]{https://www.trl.mek.dtu.dk/}
\affiliation{Department of Civil and Mechanical Engineering, Technical University of Denmark, Koppels All\'{e}, Building 403, 2800 Kongens Lyngby, Denmark}


\date{\today}

\begin{abstract}
We investigate how momentum and kinetic energy is transferred between Fourier components (the so-called triad interactions) in measured turbulent flow fields, i.e. in practical, discretely sampled signals with limited temporal and spatial domains. We empirically observe that the finite resolution in experimental investigations causes a decoupling between time and space, which broadens the phase match condition to include both spatial and temporal frequencies as predicted in Part 1. It is also empirically observed that the Fourier components may interact with a finite time delay and within a broadened frequency window (finite overlap widths) in both time and space, as compared to the usual integrals over infinite ranges where Fourier components interact by overlapping Dirac delta functions. Furthermore, it is empirically observed how the finite temporal and spatial measurement domains of velocity records can have a significant effect on the efficiency of the triad interactions and thereby on the shape and development of measured velocity power spectra. These finite spatial/temporal domains thus influence the measured spatial and temporal development of turbulence, the possibility for non-local interactions and hence also non-equilibrium turbulence, e.g. fractal grid generated turbulence. 
\end{abstract}



\maketitle 


\section{\label{sec:1}Introduction}

The analysis of triad interactions, i.e., the interactions between Fourier modes of the velocity field of a turbulent flow, is a powerful tool for the understanding of turbulence and the prediction of the dynamic development of various statistical quantities describing turbulent flows. Most often, the triad interactions are treated analytically by Fourier integrals over a three-dimensional spatial velocity field, possibly weighted by a spatial window function delimiting the spatial region of interest~\cite{Davidson,15}. The time development of the flow is then described by solving the Navier-Stokes equation in Fourier space. The Fourier integral formulation then demands phase match between the interacting modes in order to deliver efficient results of the interactions.

While theoretically, space and time are coupled at any point in the flow through the definition of velocity, $\bm{u}(\bm{r},t)=\frac{d\bm{r}}{dt}$, the uncertainty in real measurements of the exact location of the measurement point within a finite size measurement volume, means that space and time are effectively decoupled at small scales, and it will be necessary to consider time as a fourth independent stochastic variable. In Part 1~\cite{Part1}, we investigate the effect of performing a four-dimensional Fourier transform with four random variables, the three spatial coordinates and time, on a highly turbulent velocity field and we describe the resulting four-dimensional modal interactions. The result is that we see further possible mode couplings involving both spatial and temporal modes and also combinations of spatial and temporal interactions.

The conventional analysis of the three spatial phase matched Fourier modes allows the study of the basic properties of the dynamics of turbulent flows with different initial conditions and assuming different statistical modal distributions, see for instance Kraichnan~\cite{4}. However, real experimental conditions do have a great influence on the measured time development of the modal dynamics and the possible spectral modal couplings~\cite{iTi2021}, as can be observed in fractal grid generated turbulence and other non-equilibrium turbulence~\cite{25,26,ReconsideringK41}. 

In the present work, we analyze the dynamic mode couplings in turbulent flows as they appear in realistic measurements. We consider the effect on the mode couplings caused by digital sampling and due to finite measurement volume, finite time and velocity resolution as well as through the influence of finite spatial measurement regions and finite temporal record lengths. We supplement the theoretical analysis by examples from a simple, one-dimensional Navier-Stokes equation solver~\cite{2} and display results from a flow containing a single, strong oscillatory mode as well as a dual oscillatory mode experiment. These results demonstrate respectively the delays in the modal interactions in the cascade evolution and the broadening of the phase match condition in the modal interactions due to decoupling between space and time, as well as the broadening of the phase match condition for finite measuring volumes. The experimental work with shedding from dual rods that are allowed to interact can be found in more recent work~\cite{Dualpaper}. 

Laser Doppler anemometry measurements in a turbulent jet demonstrate how the finite measuring volume decouples space and time, resulting in a randomization of the the amount of time that the tracer particle resides in the measurement volume and hence the randomness in the time for the measurement. 

Throughout the analysis, we must keep in mind that the (Fourier) mode structure of a turbulent flow is a mathematical tool where the modes are defined by the spatial and temporal resolution imposed on the flow problem by us when we decide the experimental conditions and select the parameters of the mathematical analysis.

\section{Finite domains \& peak widths}

For simplicity and without loss of generality, we assume Cartesian spatial coordinates, $\bm{r}=(x,y,z)$, and time, $t$. We denote the wave vector, $\bm{k}= (k_x, k_y, k_z)$, and angular frequency, $\omega$, in Fourier space. The analysis can in a straightforward manner be adapted to other coordinate systems. Let the circumflex, `$\hat{\cdot}$', denote the Fourier transform. 

The theoretical expression for the spectral decomposition of the velocity, $\bm{u}$, in wave vector space, 
\begin{eqnarray}
\bm{\hat{u}}(\bm{k},\omega) = \iiiint\limits_{-\infty}^{\infty} \bm{u}(\bm{r},t) e^{-i(\bm{k}\cdot \bm{r}-\omega t)}\, d\bm{r} \, dt,
\end{eqnarray}
does not describe a practical experiment, since it involves continuous integrals over an infinite record in space and time. Consequentially, when applying the same analysis to the nonlinear term in the Navier-Stokes equation (as was done in equations 11-13 in Part 1\cite{Part1}), the wave vectors of the interacting waves, $\bm{k}$, $\bm{k}_1$ and $\bm{k}_2$, must match within spatial delta functions, yielding $ \bm{k} - \left ( \bm{k}_1 + \bm{k}_2 \right ) =0 $ and $ \omega - \left ( \omega_1 + \omega_2 \right )=0$ or, if Taylor's hypothesis cannot be invoked, collectively as $\left ( \bm{k} -  (\bm{k}_1 + \bm{k}_2 ) \right )\cdot \bm{r} - \left ( \omega -  (\omega_1 + \omega_2 ) \right ) t = 0$. 

In case of low turbulence intensities ($\lesssim 20\%$), we can apply Taylor's frozen field hypothesis, and the temporal and spatial phase match conditions become identical, and we could continue the analysis with the three-dimensional triad interactions\cite{Part1}, $\bm{k} -  (\bm{k}_1 + \bm{k}_2 ) = 0$.

For practical signals of finite extent and digital sampling, the situation is rather different from the ideal expressions with infinite domains and interactions between ideal delta functions. When we analyze measurements consisting of $\bm{N} = (N_x,N_y,N_z)$ digital samples along all three spatial coordinate directions, collected over finite record lengths, $\bm{L}=(L_x,L_y,L_z)$ for spatial record length and $T$ for temporal record length digitized over $M$ samples, we find that an interaction is not just limited to phase matched wave numbers, but can take place within a range of wave numbers and frequencies. 

Let $\Delta k_i = 1/L_i$ for each spatial dimension $i$ and $\Delta \omega = 1/T$ be the respective resolutions in wavenumber and frequency space, while $\Delta \lambda_i = L_i/N_i$ and $\Delta t = T/M$ are the resolutions in physical space. It is important to realize that $L_i$, $T$, $M$ and $N_i$ are chosen by us to best describe the measurement, but that especially in free flows these limits are hard to define due to the fluctuating nature of the flow boundaries. 

In practice, we can estimate $L_i$ as the maximum dimension in each direction of a box surrounding the volume of interest, for example with one of the dimensions of $L_i$ given by the half width of a round jet at the chosen distance from the exit, or as the largest eddy of interest in a particular experiment. $T$ can e.g. be estimated as the turnover time of the largest eddy of interest.

With finite rectangular windows, $W_L$ being a box-car spatial window of extent $L$ in each direction and $W_T$ being the temporal window of extent $T$, the wave vector delta functions are replaced by a product of wave number sinc-functions:
\begin{widetext}
\begin{eqnarray}
\hat{W}_L &=& \int_{-L_x/2}^{L_x/2} \int_{-L_y/2}^{L_y/2} \int_{-L_z/2}^{L_z/2}e^{-i \left ( \bm{k} -  (\bm{k}_1 + \bm{k}_2 ) \right ) \cdot \bm{r}} \, d\bm{r}   \notag  \\
&=&  L_x \, L_y \, L_z \,\mathrm{sinc} \left [ \left ( \bm{k} -  (\bm{k}_1 + \bm{k}_2) \right )_x \frac{L_x}{2} \right ]
    \mathrm{sinc} \left [ \left ( \bm{k} -  (\bm{k}_1 + \bm{k}_2) \right )_y \frac{L_y}{2} \right ]
    \mathrm{sinc} \left [ \left ( \bm{k} -  (\bm{k}_1 + \bm{k}_2) \right )_z \frac{L_z}{2} \right ] 
\end{eqnarray}
\end{widetext}
where $\left ( \bm{k} -  (\bm{k}_1 + \bm{k}_2) \right )_i$ represents the $i$'th element of the vector $\left ( \bm{k} -  (\bm{k}_1 + \bm{k}_2) \right ) $. The frequency delta function is correspondingly replaced by the sinc-function frequency window
\begin{equation}
\hat{W}_T = \int_{-T/2}^{T/2} e^{-i \left ( \omega -  ( \omega_1 + \omega_2 ) \right ) t} \, dt =
    T \mathrm{sinc} \left [ \left ( \omega - \left ( \omega_1 + \omega_2 \right ) \right ) \frac{T}{2} \right ].
\end{equation}
These integrals thus show that interactions may take place not just among the strict (delta function) phase matches $\bm{k} -  \left ( \bm{k}_1 + \bm{k}_2 \right ) =0$ and $\omega -  \left (\omega_1 + \omega_2 \right ) = 0$, but in a range of wavenumbers and frequencies defined by the width of the sinc-functions, which in turn are defined by the spatial and temporal record lengths, $L_i$ and $T$. When deriving the expression for the velocity power spectrum of a finite time record, $S_T(\omega)$, using the window functions, the analytical spectrum for infinite records, $S(f)$ is thus convolved with a sinc$^2$-function: $S_T(\omega) = S(\omega) \otimes T \mathrm{sinc}^2\left ( \omega - (\omega_1 + \omega_2 ) \frac{T}{2} \right )$. 

With the broadened phase match condition, $\left ( \bm{k} -  (\bm{k}_1 + \bm{k}_2 ) \right )\cdot \bm{r} - \left ( \omega -  (\omega_1 + \omega_2 ) \right ) t = 0$, a mismatch in the spatial phase, $\delta \bm{k} = \left ( \bm{k} -  (\bm{k}_1 + \bm{k}_2 ) \right ) \neq 0$, can be compensated by a temporal phase term, $ \delta \omega = \left ( \omega -  (\omega_1 + \omega_2 ) \right ) t \neq 0$, such that $\delta \bm{k} \cdot \bm{r} = \delta \omega t$. From $t = \delta \bm{k} \cdot \bm{r} / \delta \omega$, the range of wave numbers can be related to the frequency range and thereby to the time duration of the interaction if we know the relation between $\bm{k}$ and $\omega$, i.e. the velocity of the traveling Fourier waves, or in other words the dispersion relation for the Fourier waves. However, in a turbulent field there is no known single dispersion relation. The interesting thing about fluid structures is that they have different velocities (after all, they are velocity waves!), and the velocities of these waves are independent of the spatial frequencies. In other words, a certain spatial velocity structure may move with a velocity independent of the structure itself. This, at least, is valid for the small-scale velocity structures. 

The need to consider time a stochastic variable is often linked to the sampling process. Consider the sampling process for the laser Doppler instrument with a finite size measurement volume: The time for a sample is measured as the detector signal exceeds a certain threshold which happens when the seed particle enters the laser beam. The velocity is measured from the total particle trajectory through the measurement volume. These two events are separated by a random time depending on the exact particle trajectory. Similar considerations apply to many optical measurements and even to hot-wire measurements.

In the next sections, we shall further examine the effects of digital sampling on  the triad interaction widths.

\section{Digital triad interactions and finite interaction widths}

In practical situations, we perform measurements by digital sampling. For simplicity, we assume equidistant sampling. 

Let $\bm{L}=(L_x,L_y,L_z)$ and $T$, respectively, remain our representations of the greatest dimensions in three spatial directions and time of the flow we are interested in. Let $\pmb{\mathtt{N}} = (\mathtt{N_x},\mathtt{N_y},\mathtt{N_z})$ be the number of spatial samples in each coordinate direction and $\mathtt{M}$ the number of samples in the temporal direction. We also select the smallest resolvable scales of interest; The smallest spatial sampling interval in the three spatial coordinate directions can be expressed as $\Delta \bm{r} = \left ( \Delta r_x, \Delta r_y, \Delta r_z \right ) = \left ( \frac{L_x}{\mathtt{N_x}}, \frac{L_y}{\mathtt{N_y}}, \frac{L_z}{\mathtt{N_z}}\right )$. 
The smallest temporal sampling interval can be denoted $\Delta t = \frac{1}{f} = \frac{2 \pi}{\omega} = \frac{T}{\mathtt{M}}$ (where $f$ is the frequency). 

The digitally sampled velocity values in a Cartesian $(x,y,z)$ coordinate system, $\pmb{\mathtt{u}}=(\mathtt{u_x},\mathtt{u_y},\mathtt{u_z})$, are then given in terms of the analytical velocity $\bm{u} = \left ( u_x, u_y, u_z\right )$: 
\begin{equation}
    \pmb{\mathtt{u}} (\mathtt{n_x}, \mathtt{n_y}, \mathtt{n_z}, \mathtt{m}) = \bm{u} \left (  \mathtt{n_x} \Delta r_x, \mathtt{n_y} \Delta r_y, \mathtt{n_z} \Delta r_z, \mathtt{m} \Delta t\right )
\end{equation}
where each of the three spatial indices $\pmb{\mathtt{n}} = (\mathtt{n_x},\mathtt{n_y},\mathtt{n_z})$ can take on the numbers $\mathtt{n_i} = 0,1,\ldots,\mathtt{N_i}-1$, where $\mathtt{i}$ represents the coordinate index, and for the time index $\mathtt{m} = 0,1,\ldots,\mathtt{M}-1$. 
The window functions will be implicit in the digital Fourier transform as a consequence of the finite limits of the summations representing the three spatial coordinate directions and time, respectively. 

We proceed with the digital version of the Fourier transform of the sampled velocity values:
$$ \hat{\pmb{\mathtt{u}}} \left (\mathtt{k_x}, \mathtt{k_y} , \mathtt{k_z}, \mathtt{\omega} \right ) = \Delta r_x \Delta r_y \Delta r_z \Delta t \times$$
\begin{equation}\label{eq:55}
    \sum_{\mathtt{n_x}=0}^{\mathtt{N_x}-1} \sum_{\mathtt{n_y}=0}^{\mathtt{N_y}-1} \sum_{\mathtt{n_z}=0}^{\mathtt{N_z}-1} \sum_{\mathtt{m}=0}^{\mathtt{M}-1} \pmb{\mathtt{u}} \, e^{-i\mathtt{k_x} \mathtt{n_x} \Delta r_x} e^{-i\mathtt{k_y} \mathtt{n_y} \Delta r_y} e^{-i\mathtt{k_z} \mathtt{n_z} \Delta r_z} e^{i\omega \mathtt{m} \Delta t}
\end{equation}
where $\hat{\pmb{\mathtt{u}}} \left (\mathtt{k_x}, \mathtt{k_y} , \mathtt{k_z}, \mathtt{\omega} \right ) = \hat{\bm{u}} (\mathtt{n_x}\Delta k_x,\mathtt{n_y}\Delta k_y,\mathtt{n_z}\Delta k_z, \mathtt{m} \Delta \omega)$ is described by the digital wavenumbers $\pmb{\mathtt{k}} = \left ( \mathtt{k_x}, \mathtt{k_y} , \mathtt{k_z} \right )$ and digital angular frequency $\mathtt{\omega} = 2 \pi f = \frac{2\pi }{\Delta t} = \frac{2 \pi M}{T}$.

Using the discrete windowed Fourier transform when Taylor's hypothesis can be invoked, we arrive at the digital expression for the nonlinear term in the Navier-Stokes equation corresponding to Equation (15) in Part 1~\cite{Part1}, where $\pmb{\mathtt{k}_1}$ and $\pmb{\mathtt{k}_2}$ indicate the Fourier components interacting to produce the Fourier component with wave vector $\pmb{\mathtt{k}}$: 
\begin{equation}\label{eq:NL}
    \hat{\pmb{\mathtt{NL}}}(\mathtt{k_x}, \mathtt{k_y} , \mathtt{k_z}, \mathtt{\omega}) = \sum_{\pmb{\mathtt{k_1}}=0}^{\pmb{\mathtt{N}}-1} \sum_{\mathtt{\omega_1}=0}^{\mathtt{M}-1} 
    \sum_{\pmb{\mathtt{k_2}}=0}^{\pmb{\mathtt{N}}-1} \sum_{\mathtt{\omega_2}=0}^{\mathtt{M}-1} \left [ \left (-i \pmb{\mathtt{k}_2} \cdot \hat{\pmb{\mathtt{u}}}(\pmb{\mathtt{k}_1},\mathtt{\omega_1}) \right ) \hat{\pmb{\mathtt{u}}}(\pmb{\mathtt{k_2}},\mathtt{\omega_2}) \right ],
\end{equation}
where $\pmb{\mathtt{k}_1}$ and $\pmb{\mathtt{k}_2}$ are allowed to range freely. Equation~(\ref{eq:NL}) represents the process that must be implemented in any digital measurement of the interacting velocity modes resulting from the nonlinear term in Navier-Stokes equation. The digital sampling interval defines the resolution of the mode structure defining the measured turbulence. As in any digital measurement, we also have to consider the sampling interval in relation to the frequency content of the analog velocity trace to make sure to comply with the Nyquist sampling criterion to avoid aliasing. Equations~(\ref{eq:55}) and~(\ref{eq:NL}) have a form which will be applied in a real digital measurement.


The spectral windows resulting from the finite spatial and temporal lengths of the velocity record thus result in a finite spatial and temporal width of the interactions. This is of course well known, but in the context of the present work, we emphasize that the Fourier components of the spatial and temporal windows constitute the form of the temporal and spatial modes that interact through the nonlinear term in the Navier-Stokes equation. 

We may define the spatial width of the interaction as the width between the zero crossings of the spatial sinc-function, 
\begin{equation*}
    \Delta k_i = 4 \pi /L_i
\end{equation*}
In the same manner, the extent of the temporal mode may be defined as
\begin{equation*}
    \Delta \omega = 4 \pi /T.
\end{equation*}


In Part 1 of this two-part paper series~\cite{Part1}, we considered the modes to be mathematical plane waves making up the basis for the Fourier transform. However, in this paper, knowing the short time mean velocity of the fluid, $\bm{u}_0 (\bm{r},t)$, at the measurement point, we can connect the spatial modes to the temporal modes, albeit only for a short time. Knowing $\bm{u}_0$, we find $\omega = \bm{u}_0\cdot \bm{k}$ or $\Delta \omega = \bm{u}_0\cdot \Delta \bm{k}$. We can then estimate the time duration of the interaction by the Fourier uncertainty as $\Delta \tau \propto \frac{2\pi}{\Delta \omega} = \frac{2 \pi}{ \bm{u}_0 \cdot \Delta \bm{k}}$, expressed by the wavenumber range. We find that the duration of the interaction, $\Delta \tau$, is inversely proportional to the wavenumber span and the instantaneous wave velocity. 

As a simple example, assume $\Delta k = 10^3 \, \mathrm{m}^{-1}$ corresponding to a $1 \, \mathrm{mm}$ wavelength, moving with a velocity $u_0 = 1\, \mathrm{ms}^{-1}$ and assuming Taylor's hypothesis can be invoked, then $\Delta \tau = \frac{2\pi}{u_0 \Delta k} = \frac{2 \pi}{1 \cdot 10^3} = 6.28\cdot 10^{-3} \, \mathrm{s}$. Thus, a typical interaction lasts a few milliseconds. This agrees with the measurements in Josserand \textit{et al.}~\cite{3}. 

Thus, the measured modal interaction time, in addition to depending on the velocity magnitude, also depends on the wavenumber span and thus also on the spatial extent of the measurement region and the resolution in the wavenumber decomposition. Hence, the coherence of so-called `coherent structures' in turbulence is inherently limited by these finite interaction widths.

\section{Temporal and spatial overlap in modal interactions}

An interaction between two measured spectral components will depend both on the spatial extent of the wave overlap and on the time duration of the interaction. As we shall show below, the requirement of a good spatial overlap between the waves will enhance interactions between closely measured frequency components (local interactions), but in some cases an extended time duration of an interaction may allow also non-local interactions to play a significant role. 

In a high Reynolds number homogeneous turbulence, the spatial and temporal spectral ranges are very large and the spatial and temporal pulse widths correspondingly small. Thus, local interactions will be favored, while the possibility of interactions between different scales, non-local interactions, will be small. The efficiency, however, does not require the magnitudes of $\bm{k}_1$ and $\bm{k}_2$ to have the same size as $\bm{k}$. This can explain the phenomenon observed from direct numerical simulations, sometimes denoted ``local transfer by non-local interactions''~\cite{13}, where efficient interaction occurred even when $\bm{k}_1$ and $\bm{k}_2$ were much greater than $\bm{k}$. In special circumstances, e.g. in constant shear layers or near large vortices, the interaction may persist for a longer time, and non-local interactions may be important in those cases. 

These examples can also be useful in understanding the technologically and scientifically interesting case of how non-equilibrium turbulence develops downstream of fractal grids~\cite{25,26,ReconsideringK41}. The multiple, but sometimes widely separated, scales introduced simultaneously by the grid have a high degree of temporal overlap since they are introduced into approximately the same convection velocity. The energy exchange between components and the dynamic evolution of the spectra can thereby be further accelerated and will also depend directly on the distribution of scales (spatial overlap) introduced by the chosen fractal mesh scales and how efficiently they exchange energy. As opposed to the classical single scale injecting meshes, the fractal grids can thus produce turbulence that develops significantly faster, creating a pipeline of energy transfer across the spectrum towards the dissipation range. 

Other examples with clear deviations from the Richardson cascade are available~\cite{27,Bos,Liuetal} and an extensive review of observations of non-classical and so-called inverse cascades is presented in available work~\cite{28}.

\section{Modal interaction efficiency - effects of spatially finite flows}\label{sec:modalinteff}

The realization that real experiments take place in a limited spatial and temporal region and not in an infinite domain, as presumed in the conventional theory, has further consequences for the measured modal interactions and shape of the measured spectrum: The spectral windows will influence the efficiency of the interaction between the spectral components.

Consider as a simple example the interaction between two physical plane velocity waves limited to an experimental region $L^2$ by rectangular windows in space and $T$ in time. Those two physical waves will be represented in frequency space by a range of infinite plane waves.

Let us first consider what we may call the spatial overlap. Our model for the nonlinear interaction is a spectral decomposition where the measured Fourier components are the convolution of the Fourier transform of a harmonic oscillation at the center frequency and the Fourier transform of the signal window,
\begin{equation*}
\hat{\pmb{{u}}} \otimes \hat{\pmb{{W}}} 
\end{equation*}
where 
\begin{equation*}
    \hat{\pmb{{u}}}(\bm{k},\omega ) = \int_{-\infty}^{\infty} a_0 \cos \left ( \bm{k}_0\cdot \mathbf{x} - \omega_0 t \right )e^{-i\left ( \bm{k} \cdot \mathbf{x} - \omega t\right )} \, d\mathbf{x} \, dt
\end{equation*}
is the Fourier transform of a travelling wave with amplitude $a_0$ and velocity magnitude $u_0 = k_0 / \omega_0$.
\begin{equation*}
    \hat{\pmb{{W}}}_L(\bm{k}) = \int_{-\infty}^{\infty} \pmb{{W}}(\bm{r}) e^{-i\left ( \bm{k} \cdot \mathbf{x} \right )} \, d\mathbf{x} 
\end{equation*}
and
\begin{equation*}
    \hat{{W}}_T(\omega) = \int_{-\infty}^{\infty} {W}(t) e^{-i\left ( \omega t \right )} \, dt 
\end{equation*}
are the Fourier transforms of the spatial and temporal window functions, respectively.

\begin{figure}
\center{\includegraphics[width=0.75\columnwidth]{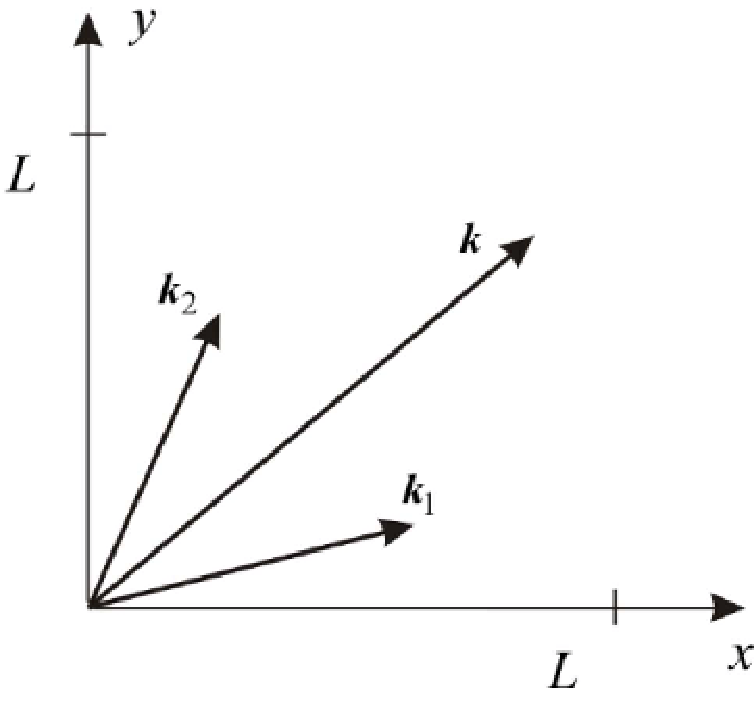}}
\caption{\label{fig:2} Wave vectors for three interacting Fourier components (plane waves) in the spatial domain $L \times L$.}
\end{figure}

\begin{figure}
\includegraphics[width=\columnwidth]{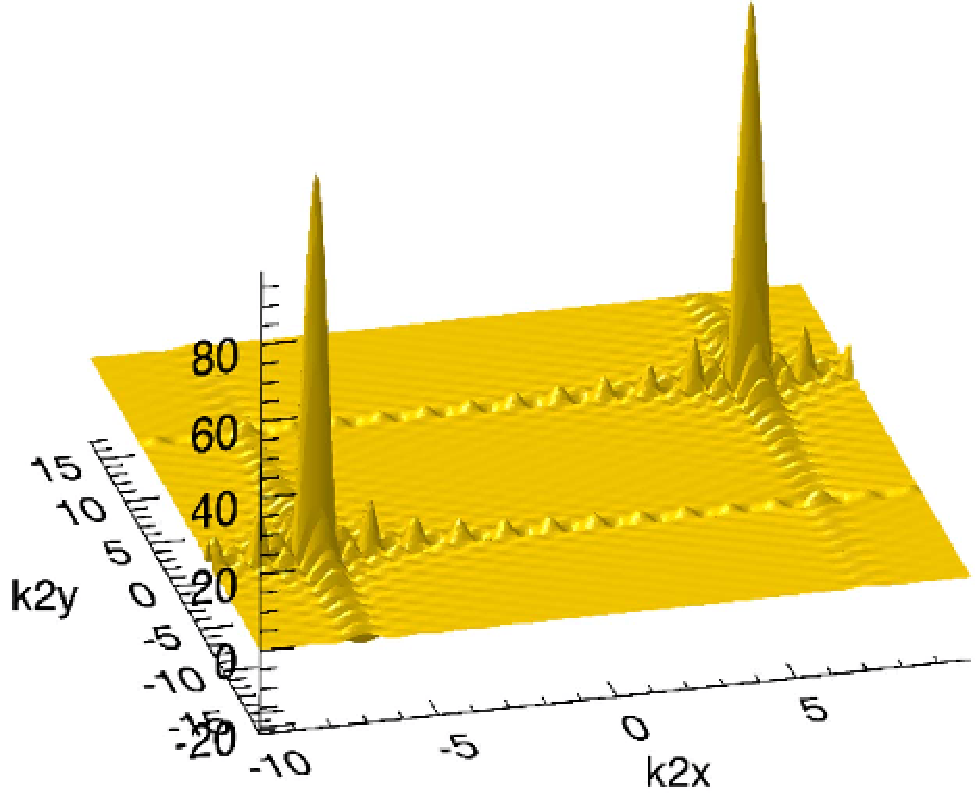}
\caption{\label{fig:3} Overlap integral as a function of $\bm{k}_2$ for a fixed value of $\bm{k}_1$.}
\end{figure}

If we pick one pair of Fourier components centered at $\bm{k}_1$ and $\bm{k}_2$, we see that they together with $\bm{k}$ define a plane as depicted in Figure~\ref{fig:2}. Through the nonlinear term, $-\left [ i\bm{k}_2\cdot \hat{\pmb{{u}}}\left ( \bm{k}_1,\omega_1 \right )\right ] \hat{\pmb{{u}}}\left ( \bm{k}_1,\omega_1 \right ) $ in Fourier space, the interaction at a point is the product of a velocity wave and the spatial derivative of a velocity wave. This product can consequently be integrated across space to find the efficiency of the interaction between the two waves. As shown in Figure~\ref{fig:3}, we can estimate the efficiency of the interaction between the two Fourier waves, $\eta_S$,  by computing the product of two modes within an area with dimension $L \times L$: 
\begin{widetext}
\begin{eqnarray}
\eta_S = \int_{-L/2}^{L/2} \int_{-L/2}^{L/2} e^{i\,\bm{k}_1 \cdot \bm{r}}e^{i\, \bm{k}_2 \cdot \bm{r}}\,dx\,dy 
   &=& \int_{-L/2}^{L/2} e^{i\left [ k_1 \cos \theta_1 + k_2 \cos \theta_2 \right ]x}\,dx \,\cdot \, \int_{-L/2}^{L/2} e^{i\left [ k_1 \sin \theta_1 + k_2 \sin \theta_2 \right ]y}\,dy \nonumber \\
   &=& L^2 \mathrm{sinc} \left [ (k_1 \cos \theta_1 + k_2 \cos \theta_2)\frac{L}{2}\right ] \, \cdot \,  \mathrm{sinc} \left [ (k_1 \sin \theta_1 + k_2 \sin \theta_2)\frac{L}{2}\right ]\nonumber
\end{eqnarray}
\end{widetext}
where $\theta_1$ and $\theta_2$ are the angles of the wave vectors to the x-axis. Within a close range defined by the sinc-functions, the waves interact efficiently. This is plotted in Figure~\ref{fig:3} with a fixed value of $\bm{k}_1$, and with $\bm{k}_2$ varying. $N$ is the number of modes or frequencies in the Fourier expansion, for example $N=200$ as used in the plot. The other values for the plot are $k_1 = |\bm{k}_1|= 10$ and $\theta_1 = \pi/3$. $k_2 =  |\bm{k}_2|$ varies from $-20$ to $20$ and $\theta_2$ varies from $0$ to $2\pi$. The peak in the fourth quadrant occurs because $k$-vectors pointing in the opposite direction have the same overlap as the ones in the first quadrant. Figure~\ref{fig:3} also shows that the interaction efficiency is not limited to the center lobe, but spreads across the side lobes over large distances from the center lobe. 

If Taylor's frozen field hypothesis applies, where the waves move at the same speed, the overlap efficiency (Figure~\ref{fig:3}) as a result also remains stationary over time. In the case that Taylor's hypothesis does not apply, e.g. in high intensity turbulence, the wave speeds will differ, and the interactions will be time limited. In this case, the pattern in Figure~\ref{fig:3} will change intermittently with time.

\begin{figure*}
\includegraphics[width=\linewidth]{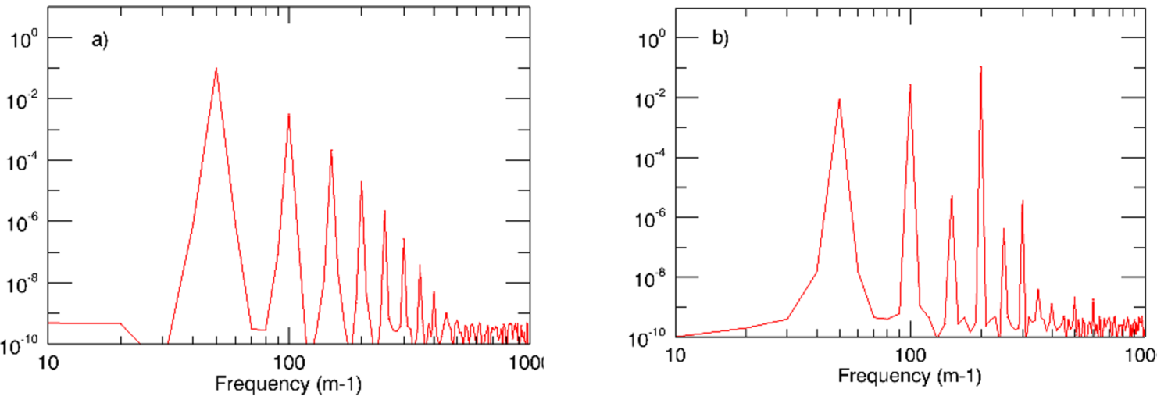}
\caption{\label{fig:4}Computer simulation of the development of the spectrum of an initial signal consisting of a single spatial frequency of $50 \, m^{-1}$. Figure 4(a): Large interaction volume, high spatial bandwidth which does not interfere with the conversion efficiency. Figure 4(b): Small interaction volume, the sinc-function spectral window influences the conversion efficiency.}
\end{figure*}

We notice that the width of the wave number uncertainty, $\Delta k_x$,  defined by the zero crossings of the sinc-function, $\Delta k_x L / 2 = 2 \pi$,  depends on the interaction area side length, $L$. The effect on the turbulence power spectrum by this spectral window is illustrated in Figure~\ref{fig:4}, which compares a computer simulation with and without the effect of the finite interaction area, see~\cite{1,2,21,iTi2021,Dualpaper} for descriptions of the simulations. The simulations are using an arbitrary scaling, hence the lack of units along ordinates of the simulation results. However, the only information of interest herein is the peak distributions across frequencies, rather than the absolute height of the peaks. Figure~\ref{fig:4} shows the development of the spectrum of an initial signal consisting of a single spatial frequency of $50\, m^{-1}$ after several cycles of triad interactions. Figure~\ref{fig:4}a presents the case of a large measurement area. The resulting high spatial bandwidth and narrow spectral band width does not cause interference between the conversion efficiency and the harmonic frequency peaks. Figure~\ref{fig:4}b shows the case of a small measurement volume. The sinc-function spectral window in this case interferes with the conversion efficiency.

\section{\label{sec:cascadeddelays}Cascaded delays -- time evolution of spectra}

As described above, energy can be passed to a mode with wave vector $\bm{k}$ by modal interactions with modes with wave vectors $\bm{k}_1$ and $\bm{k}_2$. The potential mismatch between spatial and temporal frequencies allowed to have advanced or delayed interactions to occur between velocities sampled at different times. Each interaction thus develops with a characteristic time, which depends both on the magnitude of the Fourier components (which is determined by the signal strength), the spectral bandwidth and the overlap efficiency factor, which depends on the size of the finite experimental region. 

However, the interaction time also depends on the number of time steps needed to go from $\bm{k}_1$ and $\bm{k}_2$ to $\bm{k}$. Thus, the number of delays in each modal interaction is proportional to $1+|j_1-j_2|$, where the initial value 1 accounts for the first interaction process and $j_1$ and $j_2$ are two wavenumber indices (integers) numbering the orders of the harmonics of the wave vectors interacting to produce the resulting peak in wavenumber space. 

\begin{figure*}
\includegraphics[width=\linewidth]{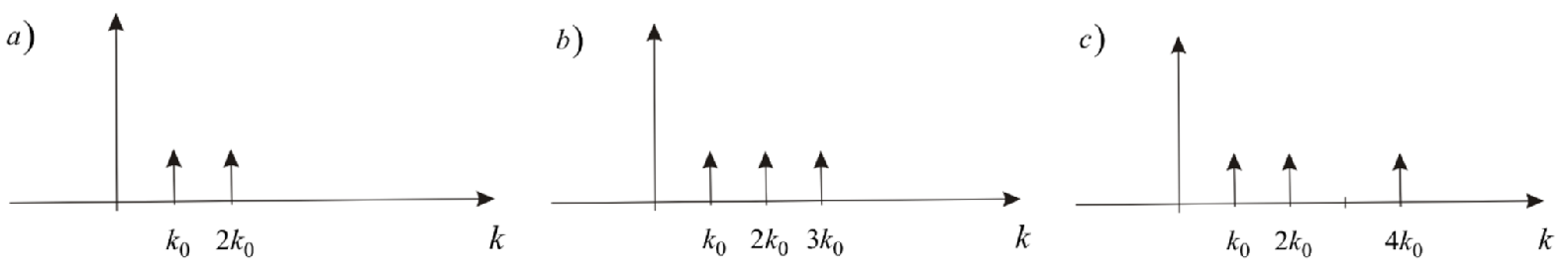}
\caption{\label{fig:5abc}(a) Frequency doubling. (b) Frequency tripling. (c) Frequency quadrupling.}
\end{figure*}

The cascaded delay in time is proportional to the sum of time delays. Let us give a couple of examples, where we consider individual modal interactions in $k$-space. We consider the presence of only one spatial input frequency component, $k_0$, and look at a few simple cascade processes for the formation of harmonics. Three different low order cascade processes are illustrated in Figure~\ref{fig:5abc}.\\

\noindent \textbf{Frequency doubling:} $k_0 \Rightarrow 2 k_0$ \\
\noindent In this example, $k_1=k_2$, hence $j_1 = j_2$ and there is only one delay, $1+|j_1-j_2|=1$.\\

\noindent \textbf{Frequency tripling:} $k_0 + 2 k_0\Rightarrow 3 k_0$ \\
\noindent Here $k_1 = k_0$ and $k_2 = 2k_0$. Thus $1+|j_1-j_2|=2$. But $k_2 = 2k_0$ must be created first, resulting in a total cascade delay of 3. \\

\noindent \textbf{Frequency quadrupling: } $2 k_0 + 2 k_0\Rightarrow 4 k_0$\\
\noindent Here $k_1 = 2 k_0$ and $k_2 = 2k_0$. $j_1=j_2$, so there is only one delay in this interaction, $1+|j_1-j_2|=1$. But $2k_0$ must be created first, resulting in a total cascade delay of 2. Thus, the fourth harmonic would be created earlier and more efficiently than the third harmonic, which can be observed in many measured power spectra (see e.g. Figure~\ref{fig:4}b).\\

\section{\label{sec:5}Experiments and numerical simulations}

Our analysis is supported by hotwire and laser Doppler anemometry experiments. The first experiment, tracing downstream harmonics development from a single injected Fourier mode using hotwire anemometry, provides empirical data to estimate the duration of individual modal interaction processes and the total delay in a process involving a number of cascade steps.

The second experiment consists of a laser Doppler anemometry measurement in the fully developed region of a turbulent jet. The finite measuring volume causes a decoupling between time and space, resulting in random and finite values of the measured transit times of the seed particles. This adds a non-zero temporal frequency contribution $\Delta \omega \, t \neq 0$ to the total phase match $\Delta \bm{k} \cdot \bm{r} - \Delta \omega \, t = \left [ \bm{k} - (\bm{k}_1 + \bm{k}_2 ) \right ] \cdot \bm{r} - \left [ \omega - \left (\omega_1 + \omega_2 \right ) \right ] t = 0$ that allows also for non-zero spatial frequency phase matches $\Delta \bm{k} \cdot \bm{r}\neq 0$.

Lastly, a dual vortex shedding experiment demonstrates how modal interactions with finite and infinitesimal measuring volumes differ and how the finite measuring volume experiment yields a broadening of the phase match condition in comparison to the classical purely spatial frequency phase match. The results demonstrate how the randomness of the measurement of time affects the measured power spectrum and the distribution of the triad interactions.

\subsection{Modal interaction delays from hotwire measurements in a laminar jet core with a single Fourier mode injection}\label{sec:vshedding}

In one study~\cite{1}, we measured the development of the velocity power spectrum using a hot-wire anemometer in the approximately laminar core of a large free jet in air into which we injected a single large mode upstream. The jet has been used as an open wind tunnel in several published studies due to the high flow quality with a low (less than 1\%) turbulence intensity in the jet core, a uniform mean streamwise velocity and the thin shear layers at the jet edges (see e.g. Figures 4 and 5 in~\cite{Dualpaper} for measured streamwise velocity mean and variance profiles).

\begin{figure*}
\center{\includegraphics[width=\linewidth]{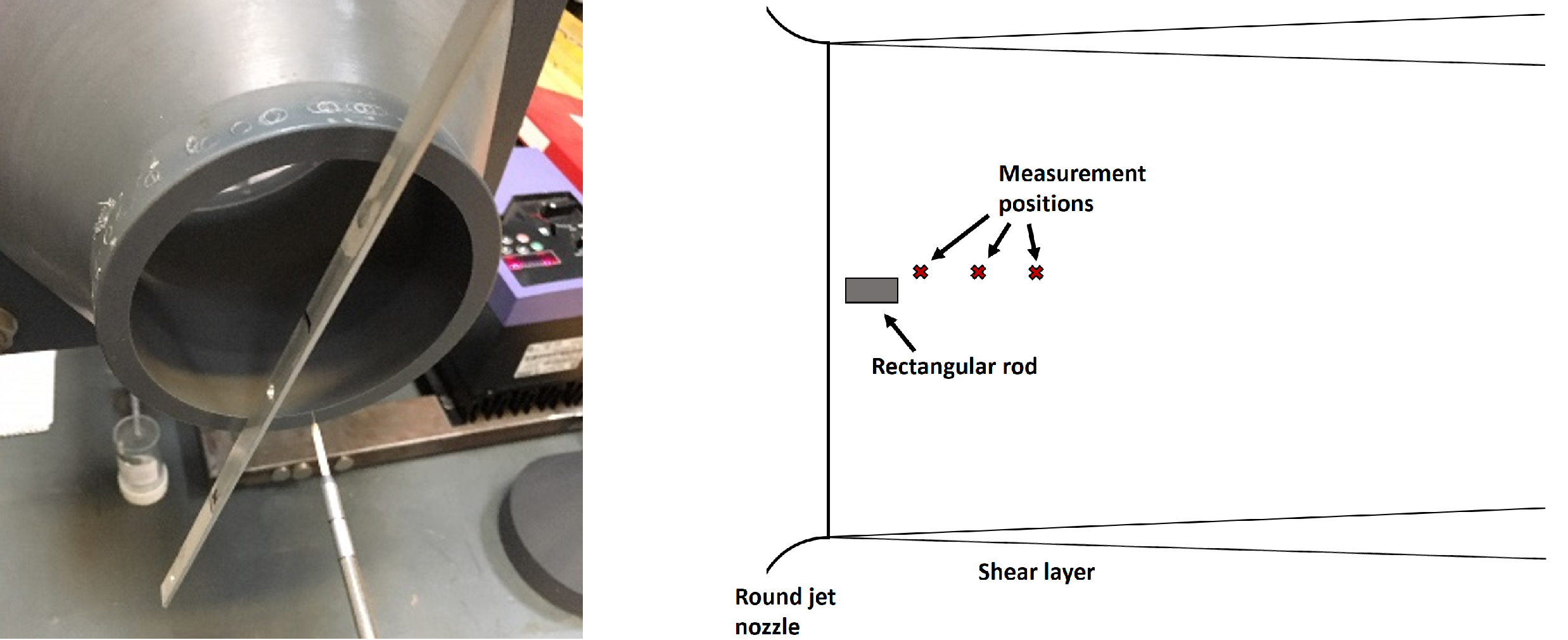}}
\caption{\label{fig:setup} (Left) Picture and (Right) sketch of the simple experimental setup used to isolate the interactions resulting from a distinct frequency shed off from a rectangular rod. The jet was large compared to the rod, such that the laminar core of the jet could create an environment similar to an open wind tunnel test section. A single component hot-wire (as seen in the left figure) was used to trace the downstream development from the initial single frequency injected by the rectangular rod.}
\end{figure*}

\begin{figure*}
\includegraphics[width=\linewidth]{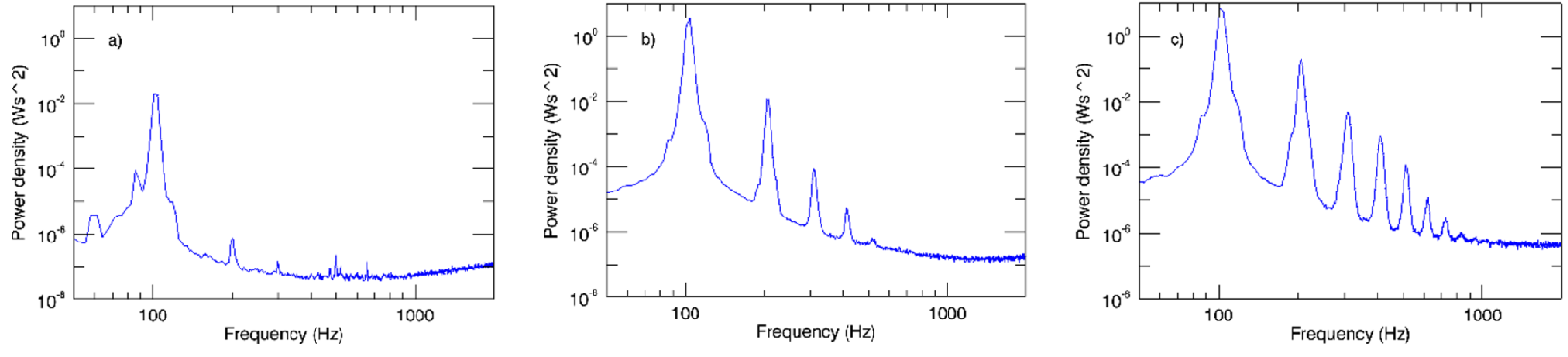}
\caption{\label{fig:5}Measured velocity power spectra with increasing downstream positions from left to right. a) $1\, mm$, b) $5\,mm$ and c) $10\, mm$ downstream.}
\end{figure*}

By measuring the power spectrum at several downstream locations from the single mode injection location, we could follow the time development of the spectrum and the appearance of the first higher harmonics generated by the nonlinear term in the Navier-Stokes equation. As can be seen in Figures~\ref{fig:setup} and~\ref{fig:5}, we used vortex shedding behind a rod with a rectangular cross-section placed across the jet to inject a narrow band oscillation which can represent a single oscillating mode close to the jet exit. A rectangular rod was chosen since the resulting shedding peak produced is sharp, much unlike the much wider shedding peak behind e.g. a circular cylinder. The jet exit had a contraction ratio of $2.4:1$ and an exit diameter of $100\,\mathrm{mm}$. The jet was run at a jet exit velocity of $1.6\,\mathrm{ms}^{-1}$. This produced a sufficiently large laminar jet core to be able to serve well as an open wind tunnel test section. The rectangular rod had a cross-section of $2 \, \mathrm{mm} \times 10\, \mathrm{mm}$, where the 2 mm width is significantly smaller than the jet exit (by a factor of 50). The rod length was $300\, \mathrm{mm}$, corresponding to three jet exit diameters and the rod was placed to cover with good margin the jet nozzle extent across the exit diameter. The rod was positioned $1 \, \mathrm{mm}$ from the nozzle and across the jet exit so that it was passing through the jet centerline. 

The sharp edges of the rectangular cross-section enabled vortex shedding with a sufficiently distinct base frequency to be able to trace the downstream development of the first and subsequent harmonics. This effect turned out to be most distinguishable in the measurements when the longest dimension ($10\,\mathrm{mm}$) of the rod cross section was aligned with the jet centerline. A single component hot-wire with a PicoScope to monitor the live spectral frequency content was used to select the optimal flow setting, measurement positions and to acquire the end result. This resulted in acquisition of the above described flow setting in ten positions along the downstream direction with $1\, \mathrm{mm}$ spacing, of which only the downstream positions $1\,\mathrm{mm}$, $5\,\mathrm{mm}$ and $10\,\mathrm{mm}$ are shown herein. The most upstream measurement point was placed $0.5\,\mathrm{mm}$ to the side away from the edge and the wake of the rod, see Figure~\ref{fig:setup}. Further detail about this experiment can be found openly available~\cite{1}.

The first spectrum, Figure~\ref{fig:5}a, shows the measured power spectrum $1\, \mathrm{mm}$ downstream from the rod where only the fundamental oscillation is present. Figure~\ref{fig:5}b shows the spectrum $5\, \mathrm{mm}$ downstream and Figure~\ref{fig:5}c the spectrum $10\, \mathrm{mm}$ downstream. 100 measured spectra have been averaged for each downstream position. The development of the cascade process is clearly visible. Knowing the velocity as a function of downstream position, we can calculate the convection time and thereby find the time history for the modal interactions~\cite{1}. In addition to the peaks, slow noise variations across frequency can be observed. These variations are introduced by the measurement probe and instrument due to electronic noise, hot-wire noise, different forms of aliasing etc. and do not represent physical spectral energy variations in the measured flow.


\begin{figure*}
\includegraphics[width=\linewidth]{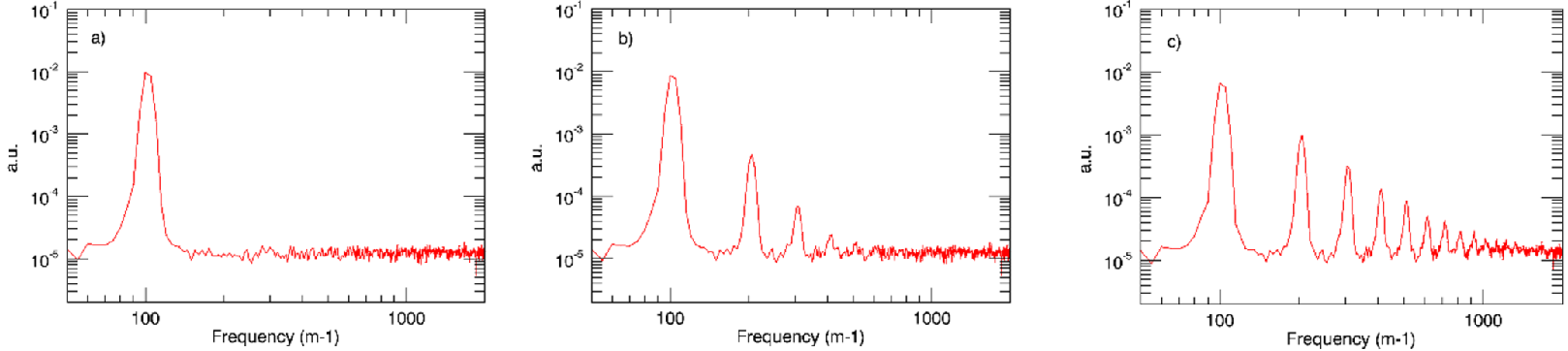}
\caption{\label{fig:6}Computed velocity power spectra at increasing time from left to right. The downstream development time has been chosen to approximately match the downstream developments displayed in Figure~\ref{fig:5}.}
\end{figure*}

In a second paper~\cite{2}, we show results of a computer calculation of a one-dimensional projection of the Navier-Stokes equation. We have implemented a one-dimensional version of the Navier-Stokes equation~\cite{21} in a program that can be executed on a laptop PC in a few minutes. By neglecting the direction of the velocity vector and only considering the velocity magnitude at a point as a function of time, the system of equations reduces to one dimension so that it is numerically solvable. The one-dimensional solution allows us to compute the kinetic energy and the turbulent structure functions directly from the Navier-Stokes equation without any additional assumptions except for the pressure term, which would require either a full three-dimensional solution for the whole flow field or a model for the pressure fluctuations~\cite{21}. We are, among many other things, able to simulate the time development of a velocity power spectrum consisting of a low intensity von K\'{a}rm\'{a}n-type turbulence to which we added a single low frequency mode.

In Figure~\ref{fig:6}, we show the development of a single Fourier mode in the same laminar jet core flow as described above and in Figure~\ref{fig:setup} and~\ref{fig:5}. Instead of a computer-generated oscillating time trace, we used the time trace actually measured near the rod as input (the same position and same velocity trace as in Figure~\ref{fig:5}a) and processed the downstream development of this time trace using our Navier-Stokes simulator~\cite{2}. 

Thus, the first plot should be identical to the measured one. (The slight differences in the initial spectra are due to the somewhat different filter settings and window functions of the PicoScope used for the measured spectra and the fast Fourier transform (FFT) settings used for the computations). The following plots show the velocity power spectrum computed by the Navier-Stokes simulator based on an average of 20 measured time traces. This should be held up against the ones measured downstream where the convection time has allowed the flow to develop. One clearly sees the effect of the cascaded delays on the development of the spectrum and the sequence of appearance of the first harmonics. This experiment has recently been extended to an analysis of the couplings between two closely spaced modes generated by two slightly different rectangular rods~\cite{Dualpaper}.

\subsection{Effect of random transit times on laser Doppler anemometry spectra in a turbulent round jet}

As a second example, we use data from a carefully conducted laser Doppler (LDA) measurement of the statistics of the self-similar region of a free, round (diameter D $=$ 10 mm) jet in air~\cite{Chunyue2022}. The Reynolds number based on the jet exit conditions is $Re \approx\,$57.000.

Sampling using the LDA is dictated by the random seeding particle arrivals in the measuring volume. More importantly for this demonstration, the finite size of the LDA measurement volume results in random transit times of the seeding particles through the measuring volume. This is dictated by several factors, such as the variation in paths that the particles can take through the measuring volume, the variations in particle velocity, optical parameters etc.. The finite size of the measuring volume thus introduces a non-zero random temporal phase factor $\Delta \omega \, t \neq 0$ into the exponent in the integral of equation (15) in Part 1, $\Delta \bm{k} \cdot \bm{r} - \Delta \omega \, t = \left [ \bm{k} - (\bm{k}_1 + \bm{k}_2 ) \right ] \cdot \bm{r} - \left [ \omega - \left (\omega_1 + \omega_2 \right ) \right ] t = 0$. Hence, even non-zero spatial phase matches can result in a zero total phase match. For a measuring volume of infinitesimal size, the temporal phase factor $\Delta \omega \, t$ would also become infinitesimal and only the spatial phase match would be required for a full phase match.

Our in-house designed optics and all-software-driven signal- and data processing ensure excellent quality of the measured data~\cite{LDAproc}. In particular, our system accurately measures and stores the transit time (residence time) of a particle track through the optical measurement volume. We can thus immediately calculate the mean and root-mean-square (RMS) residence time and get an accurate measure for the randomness of the measurement time. As an example, we have selected data at a downstream distance of 60D at the centerline of the jet. At this point, the jet mean streamwise velocity is $\sim$12\,ms$^{-1}$, the mean value of the residence time is $42\, \mu$s and the RMS value of the measured residence time is $22\, \mu$s. Thus, the fluctuation of the measurement time is in the 10 – 20 $\mu$s range. This may seem like small values, but for some applications they are not insignificant. Given a measuring volume of diameter $\sim 300\,\mu$m, if we based our velocity of the jet on the transit time, we would get about $\sim 7\,$ms$^{-1}$, but this is of course affected by a great uncertainty because of the difficulty in assessing a correct and single value for the diameter of the ellipsoidal measurement volume and for the different paths that the particles may travel through it.

From the LDA processor, the actual value of the measured velocities was obtained by spectral analysis of each individual signal burst with careful validation criteria. The data also allows us to measure the effect of randomness in the transit time measurement on the power spectrum from the decoupling between space and time due to the finite measuring volume. The power spectrum is computed from the measured velocity and the measured arrival time. The extra uncertainty in the transit time measurement will increase the noise in the spectrum, most expected in the high end of the spectrum. The power spectrum algorithm must be computed based on the measured random arrival times, and the fast Fourier transform (FFT) method cannot be used because of the random sampling times\cite{EstimationLDA}. To reduce the effect of the random sampling, the final power spectrum is averaged over 400 independent records.

\subsection{Broadening of phase match condition (time/space decoupling)}

We finally consider the effect of the randomness in the measurement time on the distribution of triad interactions. A study of these interactions offer a detailed view at the possible phase match conditions and gives unique insight into the turbulence generation. In this case, we use data from a dual vortex shedding experiment. The measurements are conducted with the same basic principle and measurement setup as described in Section~\ref{sec:vshedding}, only with two rods that introduce Fourier modes into the open wind tunnel.

\begin{figure*}
    \includegraphics[width=\linewidth]{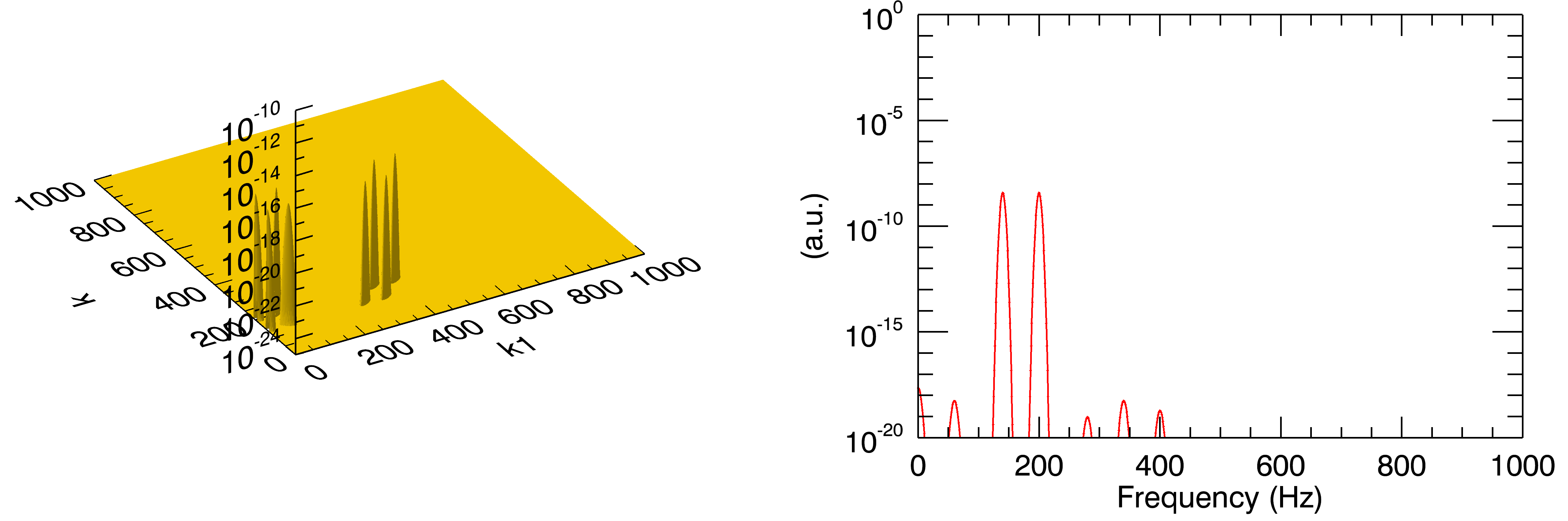}
    \caption{(Left) Purely spatial phase match results using an infinitesimal measuring volume. The interactions are dominated by the two initial frequencies and sum-, difference- and doubling-frequencies. (Right) Power spectrum at spatial frequencies $\pmb{\mathtt{k}}$. The spectrum is dominated by the initial vortex frequencies at about 140 and 200 Hz and their sum-, difference and doubling-frequencies.}
    \label{fig:Y}
\end{figure*}

Figure~\ref{fig:Y} shows the interactions (left) and the power spectrum (right) between the two Fourier modes where the temporal phase match contribution has been set to zero. In both the power spectrum and the interaction map over $\pmb{\mathtt{k}}$ and $\pmb{\mathtt{k_1}}$, it is clear that the interactions are dominated by the two initial shedding peaks (and the expected doubling, sum and difference frequencies) and hence a purely spatial phase match.

\begin{figure*}
    \includegraphics[width=\linewidth]{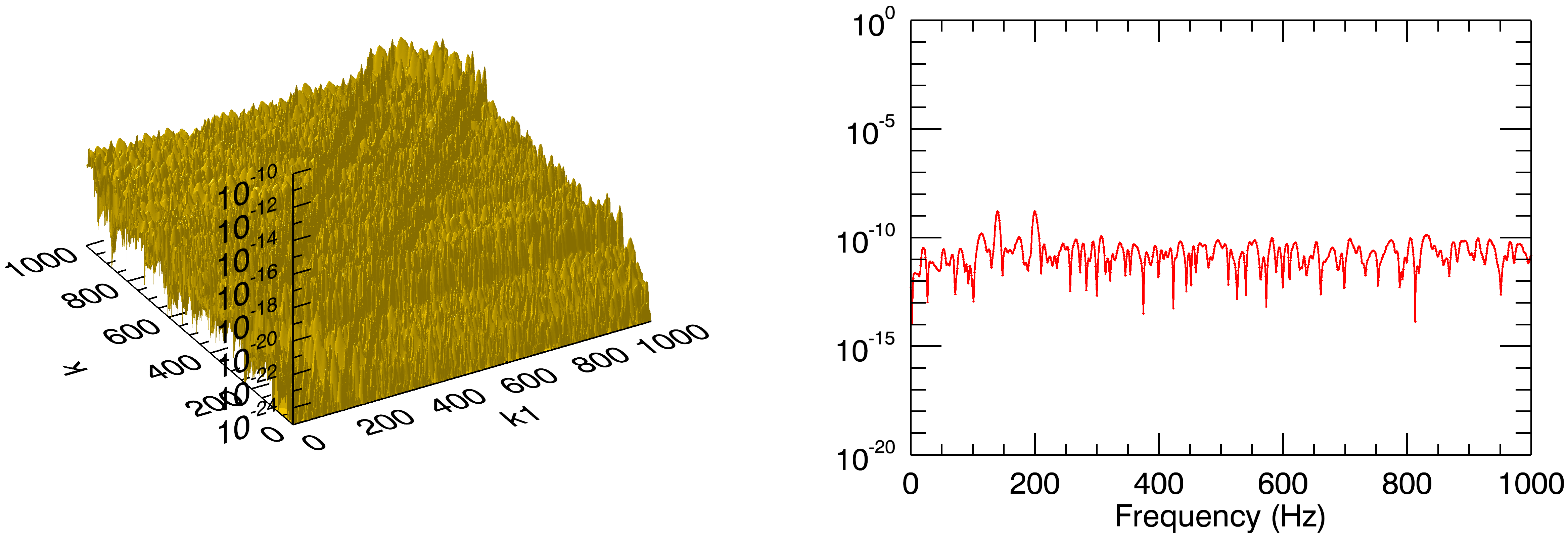}
    \caption{(Left) Spurious triad interactions at all frequencies due to random phases caused by the randomness in the measured time as well as interactions with the imposed dual modes from vortex shedding from two rectangular rods. Notice the increased level of fluctuations towards higher frequencies in accordance with the factor $\pmb{\mathtt{k}_2}$ in Equation~\ref{eq:NL}. (Right) Power spectrum at spatial frequencies $\pmb{\mathtt{k}}$. The spectrum shows a rather flat, noise-like distribution. The vortex frequencies can be discerned at about 140 and 200 Hz.}
    \label{fig:YY}
\end{figure*}

Figure~\ref{fig:YY} shows the same basic interactions as in Figure~\ref{fig:Y} (sum, difference and doublings of the 140 Hz and the 200 Hz Fourier modes) but now with less power in those modes. We now see the additional possible interactions caused by the phase jitter resulting from the uncertainty in the location of the measurement location in space and time due to the finite size of the measurement volume. As the ``real'' position and time for the measurement is totally unknown, the phase jitter contains a  broad range of frequencies resulting in the white noise level in addition to the two mode peaks apparent in Figure~\ref{fig:YY} (right) and in the enhanced fluctuations in the two diagonal lies in Figure~\ref{fig:YY} (left) caused  by interactions between the two imposed Fourier modes at 140 Hz and 200 Hz and the random phase fluctuations. Figure~\ref{fig:YY} is possibly the most direct demonstration of the additional triad interactions caused by the four-dimensional nature of the randomness of the fluctuations in the measured velocity signal.

\section{\label{sec:6}Conclusions}

The triad interactions most often analyzed assuming infinite spatial velocity fields and an analytical Fourier integral resulting in delta function phase match conditions present a too simplified description of triad interactions in the case of real measured velocity records. We analyzed how measured velocity records having finite spatial and temporal domains and finite digital sampling rates modify and expand the description of modal interactions and the development of the power spectrum of the flow.

The finite spatial and temporal domain of the velocity field creates a spectral window for the triad interactions, which introduces a spectral broadening effect and facilitates local triad interactions. Furthermore, the limited spatial domain influences the efficiency of the Fourier component overlap, which alters the shape of the generated spectrum. We illustrate the dynamic evolution of the spectra by estimating the time constant for each triad interaction and considering the sequence of the interactions needed to create a harmonic frequency.  This process illustrates the so-called spectral cascade by considering which interactions must precede the following with the associated delay. The experiments and simulations lend strong support to this picture as well.

Finally, we quoted some results from two previous papers. These examples show good agreement between a measurement in a free jet in air into which a single Fourier component is injected, for instance by a vortex shedding, and a computation of the time development of the same velocity time trace by a projection of the forces in Navier-Stokes equation onto the instantaneous velocity direction. 

LDA measurements with a finite sized measuring volume and a dual vortex shedding experiment are used to demonstrate empirically the broadening of the phase match condition for finite resolution signals.

As also discussed in the companion paper (Part 1)~\cite{Part1}, these investigations have resulted in a better understanding of triad interactions in the case of digitally sampled velocities with finite temporal and spatial measurement domains. We have also illustrated the effects of practical experiments on the processes participating in the non-linear triadic interactions and energy transfer between scales in turbulence generation and decay.

\begin{acknowledgments}
The authors wish to thank Professor Emeritus Poul Scheel Larsen for many helpful discussions. 

We also wish to acknowledge the generous support of Fabriksejer, Civilingeni{\o}r Louis Dreyer Myhrwold og hustru Janne Myhrwolds Fond (Grant Journal No. 13-M7-0039 and 15-M7-0031), and Reinholdt W. Jorck og Hustrus Fond (Grant Journal No. 13-J9-0026).

The final stages of the project would not have been possible without the support of CV by the European Research Council and of PB by the Poul Due Jensen Foundation: 

This project has received funding from the European Research Council (ERC) under the European Union's Horizon 2020 research and innovation programme (grant agreement No 803419). 

Financial support from the Poul Due Jensen Foundation (Grundfos Foundation) for this research is gratefully acknowledged.
\end{acknowledgments}



%
%

%


\bibliography{aiptemplate_NStheory}

\end{document}